\documentclass[twoside]{article}
\usepackage{graphicx}
\usepackage{fancyhdr}
\usepackage{multicol}
\usepackage{styl-ap}
\usepackage{natbib}

\begin{document}
\title{Early Super Soft Source spectra in RS\,Oph}
\author{Jan-Uwe Ness\work{1}}
\workplace{XMM-Newton Science Operations Centre, ESA, PO Box 78, 28691 Villanueva de la Ca\~nada, Madrid, Spain}
\mainauthor{juness@sciops.esa.int}
\maketitle

\begin{abstract}%
Recent Swift X-ray monitoring campaigns of novae have revealed extreme
levels of variability during the early super-soft-source (SSS) phase.
The first time this was observed was during the 2006 outburst of the
recurrent nova RS\,Oph which was also extensively covered by grating
observations with XMM-Newton and Chandra. I focus here on an XMM-Newton
observation taken on day 26.1, just before Swift confirmed the start of
the SSS phase, and a Chandra observation taken on day 39.7. The first
observation probes the evolution of the shock emission produced by the
collision of the nova ejecta with the stellar wind of the companion.
The second observation contains bright SSS emission longwards of
15\,\AA\ while at short wavelengths, the shock component can be seen
to have hardly changed. On top of the SSS continuum, additional
emission lines are clearly seen, and I show that they are much
stronger than those seen on day 26.1, indicating line pumping
caused by the SSS emission. The lightcurve on day 39.7 is highly
variable on short time scales while the long-term Swift light curve
was still variable. In 2007, we have shown that brightness variations
are followed by hardness variations, lagging behind 1000 seconds.
I show now that the hardness variations are owed to variations in
the depth of the neutral hydrogen column density of order 25\%,
particularly affecting the oxygen K-shell ionization edge at 0.5\,keV. 
\end{abstract}

\keywords{Cataclysmic variables - Recurrent novae - Spectroscopy - X-rays - individual: RS Oph}

\begin{multicols}{2}
\section{Introduction}

The 2006 outburst of the recurrent symbiotic nova RS\,Oph has attracted
a large number of observers to study the outburst in unprecedented
detail and many wavelength bands. In particular the coverage in X-rays
has considerably improved since the previous outburst in 1985 by
initiating the first high-density X-ray monitoring of a nova with the
X-ray Telescope (XRT) on board Swift, starting on day 3.38 after
initial explosion \citep{IAUC8675}.
Early shock emission originated from kinetic energy from the
nova ejecta that were dissipated in the slow, dense stellar wind of
the giant companion. The expected X-ray spectrum is that of a collisional
plasma which was confirmed by spectral models to the low-resolution
EXOSAT spectra taken in 1985 (figure 3 in \citealt{obrien92}) and the
Swift/XRT spectra \citep{bode06}. 
The X-ray grating spectrometers on board XMM-Newton (Reflection
Grating Spectrometer RGS) and Chandra (Low- and High Energy Transmission
Grating Spectrometers LETGS and HETGS) allow spectral lines to be resolved,
and in simultaneous RGS/HETGS spectra taken on day 13.8 after the initial
explosion (2006 February 12.83), bremsstrahlung continuum, H-like
and He-like emission lines, and numerous Fe lines could be identified.
A detailed analysis exploring the information from the emission
lines yielded the distribution of electron temperatures and abundances 
\citep{rsophshock}.\\

\begin{myfigure}
\hspace{-.2cm}\centerline{\resizebox{75mm}{!}{\includegraphics{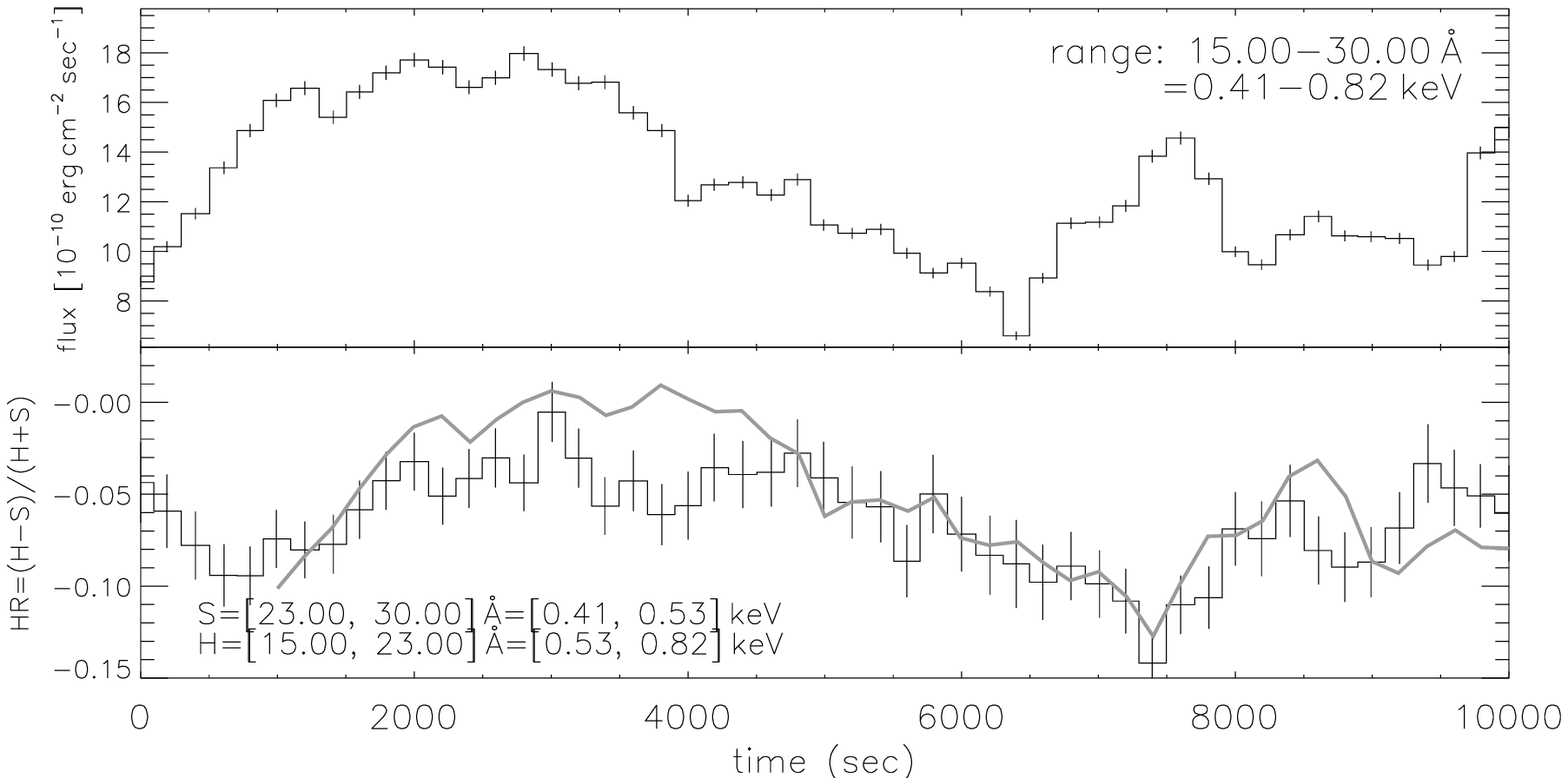}}}
\caption{Reproduction of figure~8 from \cite{ness_rsoph}. The
Chandra light curve on day 39.7, taken during the early high-amplitude
variability phase, was also highly variable on shorter time scales.
The hardness ratio, shown in the bottom, evolved with the same
variability patterns but lagged 1000 seconds behind the brightness
variations. Reproduced by permission of the AAS.}
\label{lchr}
\end{myfigure}

 Further grating observations were taken, e.g., on
days 26.1, 39.7, 54.0, 66.9, and 111.7 that were guided by continued
dense Swift monitoring (see table~1 in \citealt{rsophshock} and top
of figure~2 in \citealt{osborne11}).
During the XMM-Newton observation on day 26.1, a jump in count rate
was discovered by \cite{nelson07} that was exclusively attributed to
additional soft emission, possibly indicating the start of the
SSS phase. The bremsstrahlung continuum has faded and become softer,
and also the ratios of H-like to He-like emission line strengths
have shifted in favor of the He-like lines, indicating a cooling
trend in the shocked plasma \citep{rsophshock}.
The Swift observations of the
$\sim 60$ day SSS phase are described in \cite{osborne11}. Between
days $\sim 30$ and 45, the X-ray count rate was highly variable
between $\sim 10$ counts per second (cps) and 200\,cps, and then
stabilized at $\sim 300$\,cps (figure~2 in \citealt{osborne11}).
This was a new discovery and was later also observed in other
novae and might be a general phenomenon. During this high-amplitude
phase, a Chandra observation was taken on day 39.7, without yet
knowing about the risks of observing during times of fainter emission.
While this was the case, the soft emission was still bright enough
for a well-exposed grating spectrum that was first presented by
\cite{ness_rsoph}. The blackbody-like continuum contained deep
absorption lines from highly ionized species such as O\,{\sc viii}
and N\,{\sc vii} that were blue shifted by $\sim 1200$\,km\,s$^{-1}$.
The line profiles contained clear signs of emission lines in the
red wings (figure~5 of \citealt{ness_rsoph}) which could either
come from residual shock emission or are part of P Cyg profiles.
During this observation, the nova was also highly
variably on shorter time scales, and \cite{ness_rsoph} reported
that the hardness variations followed the same up- and down trends
but lagged 1000 seconds behind (see Fig.~\ref{lchr}).

I focus here on two new aspects:\\
1) The emission line components on top of the SSS continuum
(shown in figure~5 of \citealt{ness_rsoph}), compared to the same
lines before the SSS phase started, are much stronger than
expected from a cooling plasma (Sect.~\ref{sect:elines}).\\
2) Closer inspection of the brightness/hardness changes
during the day 39.7 observation show that the hardness changes
are due to changes in the depth of the O\,{\sc i} absorption
edge. Modeling shows that the overall column density increases
with decreasing brightness (Sect.~\ref{sect:var}).\\

\section{Observations and Analysis}

A full log of all XMM-Newton and Chandra grating observation that were
taken of the 2006 outburst of RS\,Oph is given in table~1 in
\cite{rsophshock} out of which I focus on the ones taken with
XMM-Newton between 2006 March 10, 23:04 and March 11, 02:21
(ObsID 0410180201) and with Chandra 2006 March 24, 12:25 -– 15:38
(ObsID 7296). The calibrated spectra are shown in direct comparison
in Fig.~\ref{cmplines} using the same flux units without rescaling.

\subsection{Contributions of shock emission to SSS continuum spectrum}
\label{sect:elines}

In the top panel of Fig~\ref{cmplines}, the full spectra are shown in
logarithmic units to overcome the great contrast in brightness between
the two
observations. At wavelengths $\stackrel{<}{_\sim}15$\,\AA, the two
spectra are similar in nature, mainly differing in the brightness of
the continuum. The weaker continuum on day 39.7 can be explained by
the continuation of the fading trend of
the shocks that has already been established from the observations
between days 13.8 and 26.1. The strengths of the emission lines have
hardly changed. At wavelengths $\stackrel{>}{_\sim}15$\,\AA, the 
emission lines seen on day 26.1 were outshone by the bright
continuum on day 39.7, and it is not intuitively clear how they
have evolved. In the panels below, four lines are shown in more
detail in linear units. The open histograms in the panels below
represent the day 39.7 spectrum while the colored shades are the
day 26.1 spectrum, added to the median flux of the day 39.7
spectrum, taken over the narrow wavelength range selected for
each panel. The
Mg\,{\sc xii/xi} lines have faded from day 26.1 and the
Ne\,{\sc x} line at 12.1\,\AA\ is about equally strong at both
times while these lines have become somewhat narrower.
The panels further below show that the lines on top of the SSS
continuum have become stronger from day 26.1 to 39.7, depending
on the strength of the continuum. The Fe\,{\sc xvii} line at
15\,\AA, in the Wien tail of the SSS continuum, is only
slightly stronger while the O\,{\sc viii} and N\,{\sc vii} lines,
near the peak of the SSS continuum, are much stronger, defying
the general trend of fading emission from the shocks. A similar
result has been found by \cite{schoenrich07} who show in their
figure~1 the evolution of the volume emission measure for
various emission lines as a function of their peak formation
temperature assuming collisional equilibrium. The line fluxes
of O\,{\sc vii} and O\,{\sc viii} from the observation taken on
day 39.7 yield clearly discrepant values, orders of magnitude
brighter than in observations without SSS emission, while those
emission lines that arise at shorter wavelengths are consistent
with the other observations.\\

\subsection{Spectral changes with variability during early SSS phase}
\label{sect:var}

The high-amplitude variations during the early SSS phase are still
not understood, leaving all options open such as changes in absorption,
in intrinsic brightness, temperature, or in the rate of mass loss.
A first approach to narrow down the options, I present the spectra
and how they changed with variability. During the early variability
phase of RS\,Oph, a short Chandra observation taken (day 39.7), and
the spectral evolution is illustrated in Fig.~\ref{smap_shift}. The
light curve, already shown in Fig.~\ref{lchr}, is shown in the right,
turned around by 90$^{\rm o}$ clockwise to follow the vertical time
axis in downward direction. The blue dotted line is the hardness light
curve from the bottom panel of Fig.~\ref{lchr} that was rescaled to
fit in the same graph. Along the same vertical time axis, a series
of adjacent LETGS spectra are shown in the central panel with
wavelength along the horizontal axis and flux encoded in a color
scheme, where increasing flux is represented by colors light green,
yellow, orange, red, dark blue to light blue. In the
top panel, two spectra are shown that have been integrated over the
time intervals marked with horizontal dashed lines in the central
panel shaded areas in the right panel. The colors of the dashed lines
and shades correspond to the plot style of the spectra in the top,
thus light blue and orange shadings in the right correspond to the
light blue shade and orange thick line in the top, respectively.
The two spectra have been normalized to coincide in the wavelength
range $15-22$\,\AA.\\

\begin{myfigure}
\centerline{\resizebox{80mm}{!}{\includegraphics{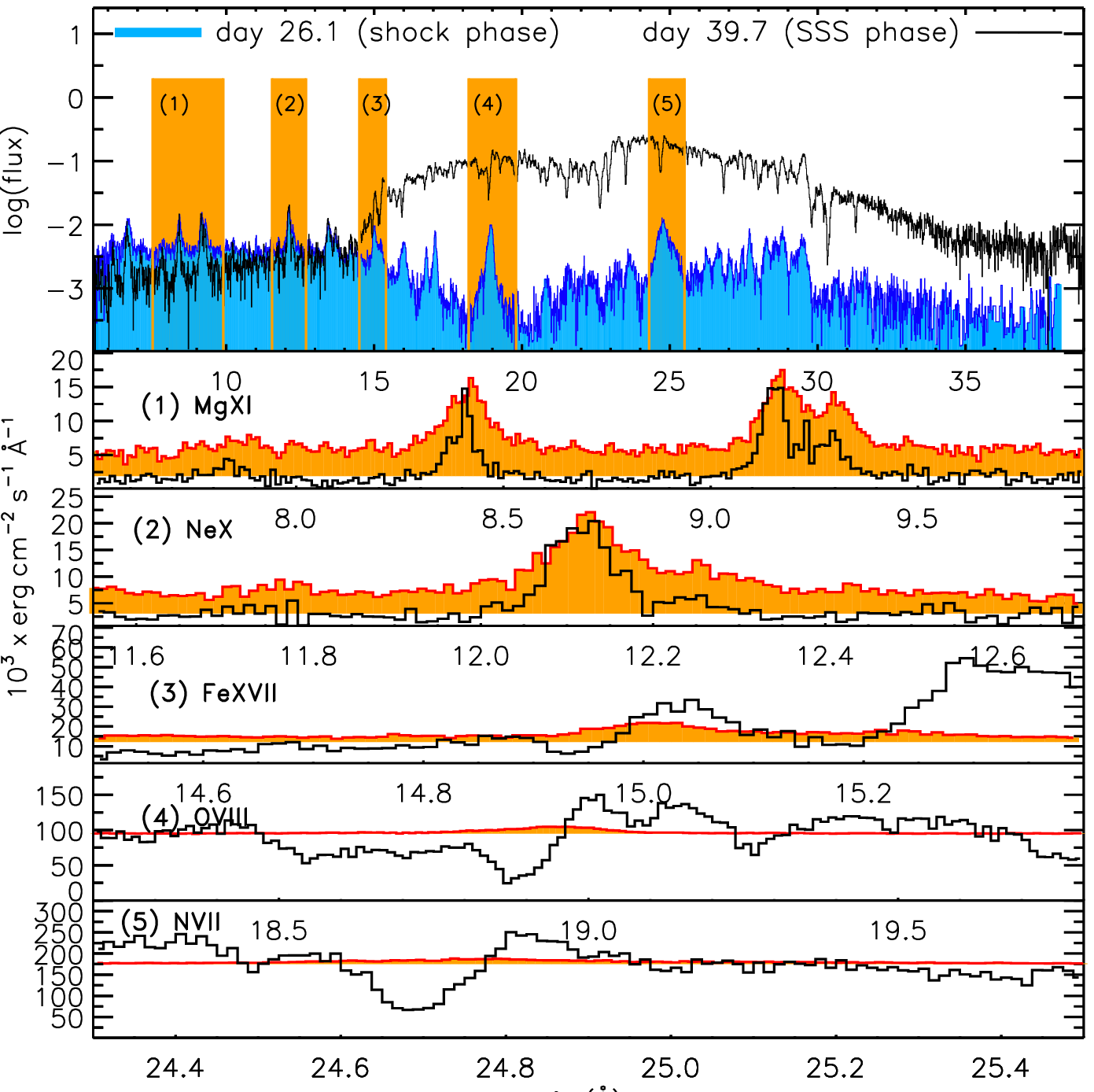}}}
\caption{Comparison of a pre-SSS spectrum of RS\,Oph taken on day
26.1 and an SSS spectrum taken on day 39.7. In the top panel the entire
wavelength range is shown in logarithmic units, illustrating that
at short wavelengths, below $\sim 15$\,\AA, the spectra are
similar, only showing a decline the Bremsstrahlung continuum
component from day 26.1 to day 39.5. In the panels below, narrow
wavelength regions are shown where the pre-SSS spectrum is added
to the median of the day 39.7 spectrum in the same flux units. While
short-wavelength lines have hardly changed, the excess emission lines
on top of the SSS continuum are much stronger than before the start
of the SSS phase.
}
\label{cmplines}
\end{myfigure}

\cite{ness_rsoph} had extracted spectra from time intervals of
bright and faint episodes, roughly corresponding to the time
intervals $0.3-1.0$\,hours of elapsed time (bright) and $1.1-2$\,hours,
respectively, and found differences in the spectral shape. I have
now chosen time intervals that are shifted by 1000\,sec in order to
probe the hardness light curve rather than the intensity light
curve. The two normalized spectra in the top clearly show that the
variability in hardness is purely owed to changes in the depth
of the O\,{\sc i} edge at 22.8\,\AA, yielding the steeper edge
1000\,sec after a low-flux episode. \cite{ness_rsoph} had found
changes in O\,{\sc i} on longer time scales between the grating
observations taken on days 39.7, 54.1, and 66.9 and explained
these by changes in the degree of ionization of oxygen. Only
neutral oxygen produces the deep edge at 22.8\,\AA\ while ionized
oxygen is more transparent to X-rays, leading to a flatter edge.
The intense soft continuum X-ray source can ionize the neutral
elements in the surroundings and thus make them more transparent.
On the other hand, if the continuum source fades, the surrounding
material recombines, leading to an increase in the O\,{\sc i}
edge.\\

\begin{myfigure}
\resizebox{\hsize}{!}{\includegraphics{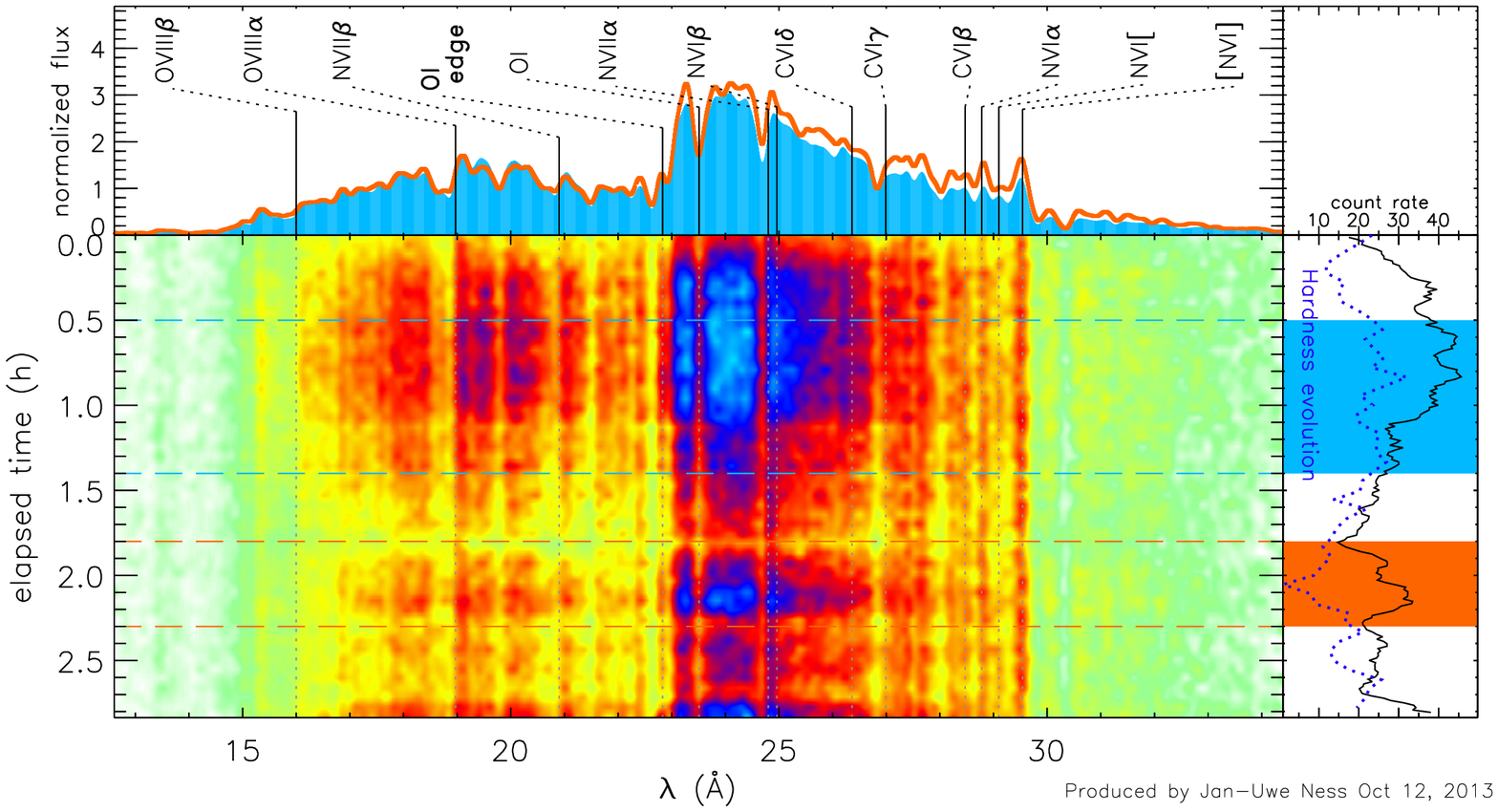}}
\caption{\label{smap_shift}Dynamic spectrum for the Chandra
observation of RS\,Oph taken on day 39.7 after the 2006 outburst.
The zero-th order light curve is shown to the right with three colored
shades marking time intervals from which the spectra in the top have
been extracted. Line labels that belong to strong transitions
are included. The central image is a brightness spectral/time map
that uses a color code from light green to light blue representing
increasing flux values.
}
\end{myfigure}

The changes in the depth of the O\,{\sc i} absorption edge could be
caused by changes in the degree of ionization of circumstellar
oxygen. \cite{ness_rsoph} argued that the observed time lag of
1000 seconds is consistent with the time scale for
ionization/recombination in a plasma with density
$\sim 10^{11}$\,cm$^{-3}$. Since all circumstellar material
would experience the same changes in their degree of ionization,
this would lead to an overall change in $N_{\rm H}$.
Another possibility is a non-uniform radial distribution of
the oxygen abundance within the ejecta that could be caused
by beta decay of oxygen isotopes produced during CNO burning.
If during times of brigher and fainter emission, different
regions of the photosphere are visible, the difference in
the oxygen abundance would only affect the depth of the
O\,{\sc i} absorption edge.\\

\begin{myfigure}
\resizebox{\hsize}{!}{\includegraphics{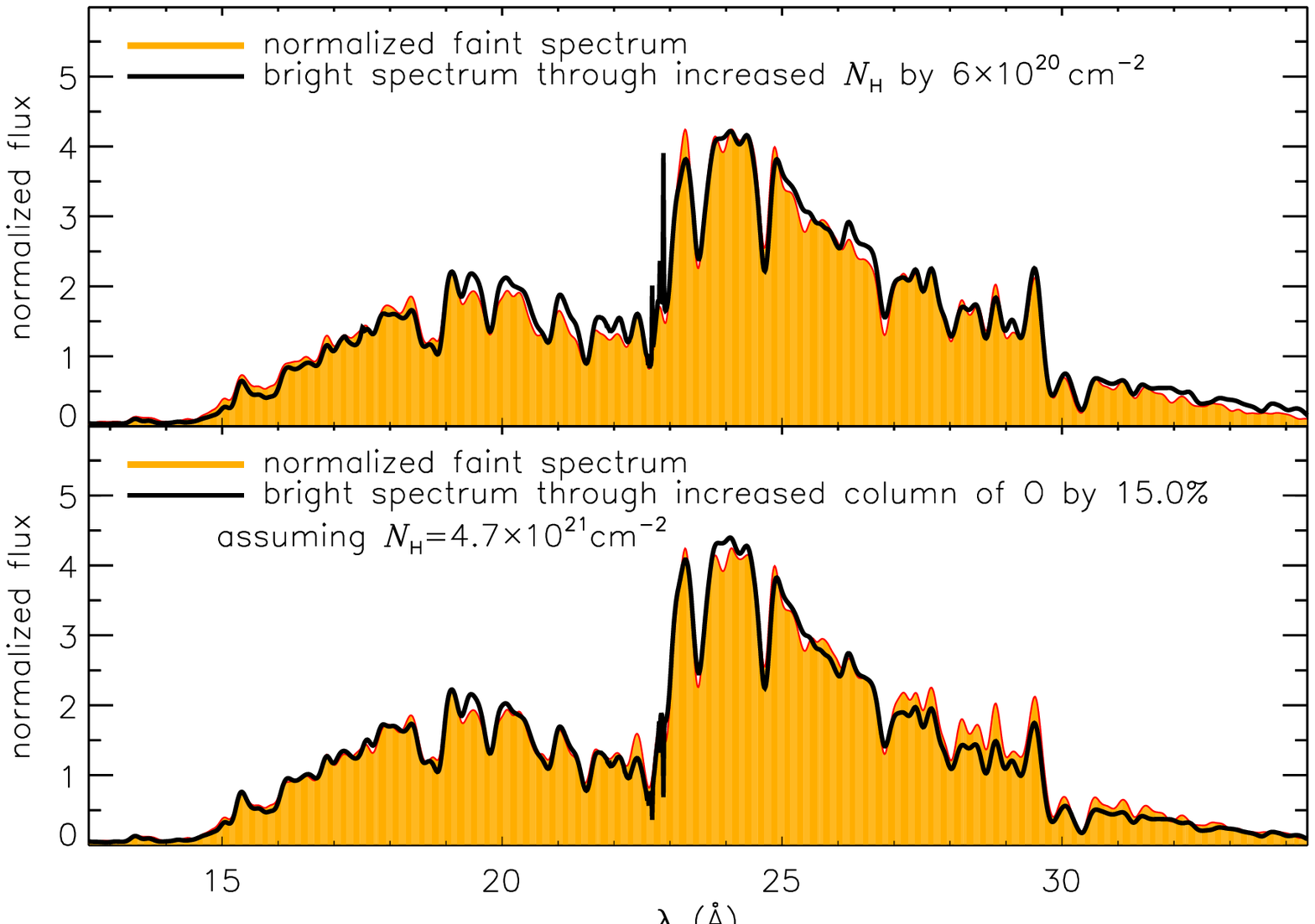}}
\caption{\label{nhfit}Testing two hypotheses to explain the spectral
changes between hard/bright (blue shades in top panel of Fig.~\ref{smap_shift})
and soft/faint (orange thick line in Fig.~\ref{smap_shift}) spectra
by the black thin lines. The normalized soft/faint spectrum is shown
as orange shadings while the black line corresponds to two types
of modifications of the bright/faint spectrum, also normalized
(see text).
{\bf Top}: Overall increase of $N_{\rm H}$ by 26\% of
$N_{\rm H,CS}$. {\bf Bottom}: Increase only of O\,{\sc i}
column by changes in O abundance of 15\%.}
\end{myfigure}

In Fig.~\ref{nhfit}, I test these two scenarios in the top
and bottom panels, respectively. Shown is the normalized
faint/soft spectrum in orange shadings (orange line in the
top panel of Fig.~\ref{smap_shift}) while the black curve
represents a modification of the brighter/harder spectrum,
trying to reproduce the faint/soft spectrum. In the top panel
of Fig.~\ref{nhfit}, the black curve was computed by dividing
the bright/hard spectrum itself by the transmission
coefficients calculated from the warm absorption model
developed by \cite{wilms00}, and then normalized. The
faint/soft spectrum is qualitatively well reproduced assuming
a value of $N_{\rm H}=6\times 10^{20}$\,cm$^{-2}$, indicating that
the change in hardness can be explained by an increase of
$N_{\rm H}$ by this amount. The full amount of the hydrogen
column was found by \cite{ness_rsoph} as
$N_{\rm H}= N_{\rm H, ISM}+ N_{\rm H,CS}=4.7\times 10^{21}$\,cm$^{-2}$,
consisting of the well-known interstellar component
$N_{\rm H, ISM}=2.4\times 10^{21}$\,cm$^{-2}$ and an additional
circumstellar component $N_{\rm H,CS}$. An increase by
$6\times 10^{20}$\,cm$^{-2}$ thus corresponds to
$\sim 26$\% of $N_{\rm H,CS}$.\\

In the bottom panel of Fig.~\ref{nhfit}, the black curve was
computed by first dividing the bright-hard spectrum by the transmission
coefficients of an absorber with $N_{\rm H}= 4.7\times 10^{21}$\,cm$^{-2}$,
assuming cosmic abundances and then multiplying by the coefficients
resulting from the same absorption model assuming a modified oxygen
abundance. The faint/soft spectrum is well reproduced up to
$\sim 27$\,\AA, but the black curve drops well below the
faint/soft spectrum at longer wavelengths. The better agreement
between the black line and the faint/soft spectrum in the
upper panel demonstrates that an overall increase of $N_{\rm H}$
has occurred rather than a change in the oxygen abundance alone.


\section{Summary and Conclusions}

A comparison of grating X-ray spectra before and after the start
of the SSS phase shows that the emission lines that were seen on
top of the SSS continuum are not simply a continuation of the early
shock emission. Only emission lines that arise at wavelengths were
strong continuum emission is present are amplified compared to the
pre-SSS level, strongly suggesting photoexcitation effects. The
line profiles in the early SSS spectrum of RS\,Oph can thus be
understood as P Cyg profiles.\\

The early high-amplitude variability in RS\,Oph and other novae
still awaits an explanation. The Chandra observation during a
minimum of these variations revealed that variability also
occurs on shorter time scales. The comparisons in this
article show that the hardness changes following brightness
changes are consistent with variations in $N_{\rm H}$ of
$\sim 26$\% which can be explained by variations in the
overall degree of ionization caused by the changes in intensity
and thus variations in the effectiveness of photoionization.


\bibliographystyle{aa}
\bibliography{cn,astron,jn,rsoph}



\end{multicols}
\end{document}